\documentclass[runningheads]{svmult}

\usepackage{makeidx}   
\usepackage{graphicx}  
\usepackage{subeqnar}  
\usepackage{multicol}  
\usepackage{physprbb}  
\makeindex             

\newcommand{\gtrsim}{\mathrel{\hbox{\rlap{\hbox{\lower4pt\hbox{$\sim$}}}\hbox{$>$}}}}
\newcommand{\lesssim}{\mathrel{\hbox{\rlap{\hbox{\lower4pt\hbox{$\sim$}}}\hbox{$<$}}}}

\begin{document}
\title*{Planetary Nebulae as Tracers of the Intergalactic  
\protect\newline Stellar Background: a Population Synthesis Theoretical Approach}
\toctitle{Planetary Nebulae as Tracers of the Intergalactic  
\protect\newline Stellar Background: a Population Synthesis Theoretical Approach}
%
%
\titlerunning{Planetary Nebulae as Tracers of the Intergalactic  
Stellar Background}
%
\author{Alberto Buzzoni\inst{1}
\and Magda Arnaboldi\inst{2}}
\authorrunning{A.\ Buzzoni \& M.\ Arnaboldi}
%
%
\institute{INAF - Osservatorio Astronomico di Bologna,
Via Ranzani 1, 40127 Bologna (Italy)
\and INAF - Osservatorio Astronomico di Torino,
Via Osservatorio 20, 10025 Pino Torinese (Italy)}

\maketitle              

\begin{abstract}
We wish to assess the relationship between the population of planetary
nebulae (PNe) and a given parent stellar population from a theoretical point
of view. Our results rely on original population synthesis models used to
estimate the expected luminosity-specific PN density accounting for different 
evolutionary scenarios and star formation histories, as observed in galaxies 
in the near Universe.  For a complete PN sample, we find that
1~PN/$1.5\,10^6$~L$_\odot$ is a safe (IMF-independent) lower limit to the
traced global bolometric luminosity of the parent stellar population.
A tentative application to Virgo cluster data allows us to place 
a lower limit at $\sim 7$\% for the global B luminosity of the cluster provided
by ``loose'' intergalactic stars.
\end{abstract}

\section{Introduction}
Recent works on diffuse light in nearby clusters have renewed interest
in planetary nebulae (PNe) as tracers of the parent stellar population.
The efficient detection of these nebulae in the field, even at 70 kpc
or more beyond the optical radius of bright galaxies (see e.g.\ Peng
et. al 2004), makes them ideal candidates to trace the ``loose''
stellar population possibly filling the ``empty'' space among
galaxies. PNe can be detected up to in nearby clusters at a distance
of 20~Mpc, like for instance for the Virgo or the Fornax cluster
(see Arnaboldi et al. 2003; Feldmeier et al. 2003).

A major issue in this regard concerns the link between PN number
density and global (bolometric) luminosity of their parent stellar population.
This ratio (the so-called ``luminosity-specific PN density'' or, 
shortly, the ``$\alpha$ ratio'', as defined first by Jacoby 1980) 
provides in fact direct information on the distinctive physical properties
of the underlying stellar population as far as its PN component can be 
picked up by the observations.

Population synthesis models can be especially useful in this sense, 
allowing us to tackle the problem from a fully theoretical point of view, thus
overcoming most of the natural limits of the empirical calibrations
(i.e.\ observational bias, incomplete sampling etc.).
On this line, we want to give here a brief progress report of our project 
aimed at calibrating ``$\alpha$'' for different evolutionary scenarios and
star formation histories as observed in the local galaxy sample and
likely consistent with the expected properties of the diffuse stellar 
component of the intergalactic medium.

\begin{figure}[!t]
\begin{center}
\includegraphics[width=0.7\textwidth, clip=]{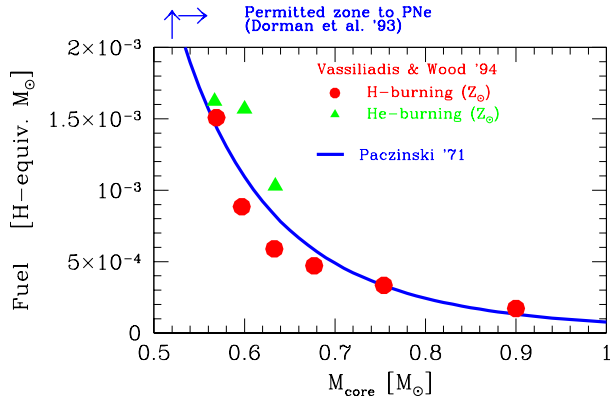}
\end{center}
\vskip -5 mm 
\caption{Theoretical fuel consumption for stars along the Hot-PAGB
evolution according to the models of Paczynski (1971; solid curve) and
Vassiliadis and Wood (1994; solid dots and triangles). Fuel is
expressed in Hydrogen-equivalent solar mass.  The stellar core-mass
range to produce PNe is marked, after Dorman et al.\ (1993).}
\label{fuel}
\end{figure}

\section{Theoretical fundamentals}

Theory of simple stellar populations (SSPs), as developed in Renzini
\& Buzzoni (1986) and Buzzoni (1989) (hereafter RB86 and B89, respectively),
provides the basic reference framework for our task, allowing us to
evaluate the value of ``$\alpha$'' for coeval and chemically homogenous
stellar aggregates.

Briefly, the expected number ($N_{\rm PN}$) of PNe for a SSP of total
(bolometric) luminosity $L_{tot}$ is:
\begin{equation}
N_{\rm PN} = {\cal B}\, L_{\rm tot}\, \tau_{\rm PN}.
\label{eq:nn}
\end{equation}
In the equation, $\cal B$ is the so-called ``specific evolutionary flux''
(a nearly age- and IMF-independent parameter of the order of 
$2\,10^{-11}~~[L_\odot^{-1}{\rm yr}^{-1}]$, see B89),
and $\tau_{\rm PN}$ is the PN lifetime (i.e.\ the time
for the nebula to be detectable in [O\,\textsc{iii}] and/or H$\alpha$
surveys). 

The PN lifetime directly modulates with the stellar core-mass
evolution, the latter depending on the nuclear fuel burnt at high
temperature ($T_{\rm eff} \simeq 10^5$~K) along the Hot Post-AGB
(H-PAGB) phase (this is a direct result of the so-called ``Fuel
Consumption Theorem'' of RB86). Figure~\ref{fuel} summarizes the model
predictions, after Paczynski (1971) and Vassiliadis and Wood (1994),
displaying the H-PAGB fuel consumption for stars of different core
masses.\footnote{Note that {\it actual} PAGB core mass of stars can
sensibly be lower than the {\it original} stellar mass when leaving
the Main Sequence Turn Off point. This is a consequence of the mass
loss processes that heavily affect stellar evolution along the red and
asymptotic giant branch phases.} In the plot, we express the fuel in
terms of Hydrogen-equivalent solar mass (i.e. 1\,g of H-equivalent
mass~= $6~10^{18}$~ergs; cf.\ RB86).

\begin{figure}[!t]
\begin{center}
\includegraphics[width=0.6\textwidth, clip=]{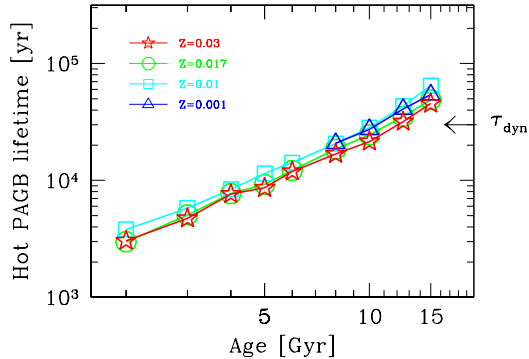}
\end{center}
\vskip -5mm
\caption{Hot-PAGB core mass lifetime, for SSP evolution according to
B89 synthesis models. A Salpeter IMF is assumed, and a Reimers mass
loss rate coefficient $\eta = 0.3$. The explored metallicity spans the
range from $Z \sim 1/20 \to 2~Z_\odot$, as labeled in the plot.
The PN dynamical timescale ($\tau_{\rm dyn}$) is also marked, for reference.}
\label{lifetime}
\end{figure}

The inferred H-PAGB lifetime for SSPs of different age, according to
B89, is reported in Fig.~\ref{lifetime}.  It can be defined 
as $\tau_{\rm HPAGB} \simeq {\rm Fuel} / \ell_{\rm H-PAGB}$,
being $\ell_{\rm H-PAGB}$ the luminosity of the
stellar core at the onset of the nebula ejection (see e.g.\ Paczynski
1971 for details). It is important to note that,
in general, $\tau_{\rm HPAGB}$ is shorter than the dynamical
timescale for nebula evaporation ($\tau_{\rm dyn} \simeq 30\,000$~yr;
cf.\ Peimbert 1990); this implies that, at least for SSPs of young and 
intermediate age,  $\tau_{\rm PN}$ should actually coincide with 
$\tau_{\rm HPAGB}$. Only for $t_{\rm SSP} \gtrsim 12$~Gyr we have
that $\tau_{\rm PN} \simeq \tau_{\rm dyn}$, so that the luminosity-specific 
PN density could eventually be written as
\begin{equation}
\alpha = {{N_{\rm PN}}\over {L_{\rm SSP}}} = {\cal B}\, \tau_{\rm PN} = 
{\cal B}\, {\rm inf} \{ \tau_{\rm HPAGB}, \tau_{\rm dyn} \}
\label{eq:alfa}
\end{equation}

\begin{figure}[!t]
\begin{center}
\includegraphics[width=0.6\textwidth, clip=]{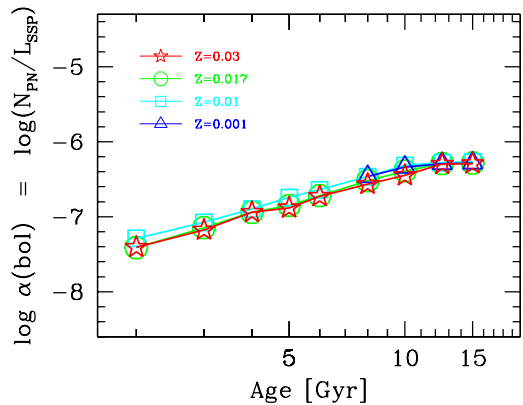}
\end{center}
\vskip -5mm
\caption{Theoretical luminosity-specific PN density for SSPs of
different metallicity. Models are according to B89, with the distinctive 
parameters as in Fig.~\ref{lifetime}.}
\label{alfa_ssp}
\end{figure}

From eq.~(\ref{eq:alfa}), and taking the B89 synthesis code as a
theoretical reference, in Fig.~\ref{alfa_ssp} we report the expected
luminosity-specific PN density for SSPs of different metallicity.
Some physical age constraints exist for SSPs to produce PNe; from one
hand, for $t_{\rm SSP} \lesssim 10^8$~yr SSP evolution is dominated by
stars with M$_* \gtrsim 5$~M$_\odot$ ending up as Supernovae. On the
other hand, no PNe are expected to form from stars with actual mass
M$_* \lesssim 0.52$~M$_\odot$ (Dorman et al.  1993; see
Fig.~\ref{fuel}); this is likely to occur at some point along SSP
evolution for ages between $10 \lesssim t_{\rm SSP} \lesssim 18$~Gyr,
the exact details depending on the efficiency of the mass loss
processes\footnote{The B89 SSP models parameterize mass loss via the
coefficient $\eta$ according to the Reimers (1975) standard
notation. A value of $\eta = 0.3$ is adopted here, as an indicative
value for the observed mass loss in the Galactic globular clusters
(see B89 for details).}.

\begin{figure}[!t]
\begin{center}
\includegraphics[width=0.74\textwidth]{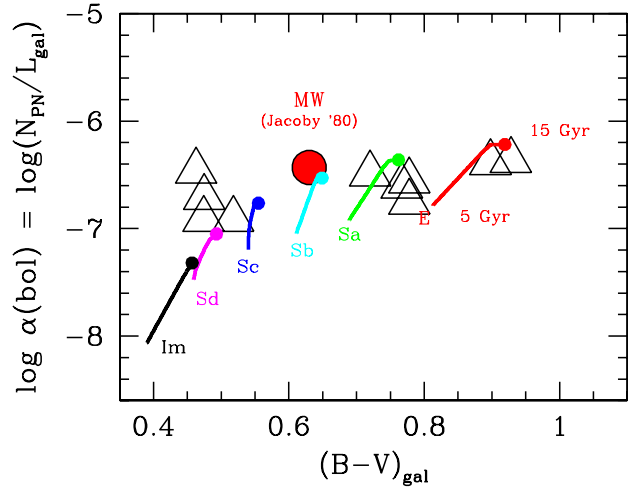}
\end{center}
\vskip -5mm
\caption{Expected luminosity-specific PN density for template galaxy
models (labeled with the corresponding Hubble type), according to
Buzzoni (2002).  Each curve tracks galaxy evolution from 5 Gyr to 15
Gyr (small solid dots), as labeled for the case of Ellipticals. Big
triangles report the empirical estimate of $\alpha = N_{\rm PN}/L_{\rm
gal}$ from the PN census of Jacoby (1980) for the galaxies in the
Local Group. The case of the Milky Way is singled out in the plot by
the big solid dot.}
\label{templates}
\end{figure}

\section{The luminosity-specific PN density in the galaxies:
  implications for the intergalactic medium}

The theoretical estimate of the luminosity-specific PN density is
important when we infer total bolometric luminosity of their parent
stellar population in the intergalactic medium. So far, we have relied 
upon the empirical calibration of $\alpha$ on the 
bulge of M31, with all the uncertainties induced by the poorly
understood effects of metallicity, age and star formation history.

Our new results, based on this improved theoretical approach, indicate
that a luminosity-specific PN density 
\begin{equation}
\alpha = 1~{\rm PN}/1.5\,10^6~{\rm L}_\odot
\label{eq:aaa}
\end{equation}
is a safe estimate for old SSPs. Quite interestingly, 
this value does {\it not} strongly depend on metallicity and, even 
more importantly, on the stellar IMF (as $\cal B$, in eq.~(\ref{eq:alfa}),
is nearly insensitive to the IMF details, as shown by RB86).  

On the basis of the derived SSP evolution, and matching the
Buzzoni (2002) template galaxy models, we have further elaborated our
synthesis scheme computing in Fig.~\ref{templates} the expected
evolution of $\alpha$ for the different star formation histories of
galaxies along the Hubble morphological sequence.
A comparison with the observed PN population for the Local Group members
(Jacoby 1980, including the relevant case of the Milky Way) 
is also reported in the figure.

As a tentative application of our results, we could match the recent 
observations of the Virgo cluster by Arnaboldi et al. (2003) in order to
assess the luminosity contribution from the intergalactic 
stellar background.  Some 37 PN candidates have been detected by Arnaboldi 
et al.\ ($\sim 25$\% of which possibly consisting of spurious Ly$\alpha$ 
emitters at $z = 3.14$, as claimed by the authors) across an area of 
0.196~deg$^2$ on the cluster. As the {\it bona fide} PNe only span the 
brightest magnitude bin of the standard luminosity function
we had to correct for sample incompleteness
according to Ciardullo et al.\ (1989).

By adopting the reference range of $\alpha$ for 15 Gyr galaxy models of Fig.~\ref{templates},
an intergalactic surface brightness (in bolometric) could be derived as
$26.0 \lesssim \mu_{\rm bol} \lesssim 28.5$~mag~arcsec$^{-2}$, 
depending on the assumed star formation history for the PN parent stellar 
population.\footnote{For our calculations
we adopt a distance modulus $(m-M) = 30.8$~mag for the Virgo cluster, 
and an absolute bolometric magnitude $M_{{\rm bol},\odot} = 4.72$ for the Sun.}
This implies a B surface brightness for intergalactic stars in the range  
$27.3 \lesssim \mu_{\rm B} \lesssim 30.3$,~mag~arcsec$^{-2}$.
Compared to the smooth B surface brightness distribution of the bright-galaxy
component of Virgo (i.e.\ $\mu_{\rm B} \simeq 27.4$~mag~arcsec$^{-2}$, see Arnaboldi et al. 2003),
this allows us to place a confident lower limit to 7\% for the 
cluster global luminosity to be provided by ``loose'' intergalactic stars.

%

\end{document}